\documentclass[secnumarabic,amssymb, amsmath,
nofootinbib,tightenlines, nobibnotes, aps, prl]{revtex4}
\usepackage{graphicx}
\begin{document}
\newcommand{\be}{\begin{equation}}
\newcommand{\ee}{\end{equation}}
\newcommand{\ba}{\begin{eqnarray}}
\newcommand{\ea}{\end{eqnarray}}

\bigskip
\vspace{2cm}
\title{CP violation in semileptonic tau lepton decays}
\vskip 6ex
\author{D. Delepine}
\email{delepine@fisica.ugto.mx}
\affiliation{Instituto de F\'{\i}sica, Universidad de Guanajuato \\
Loma del Bosque \# 103, Lomas del Campestre, \\
37150 Le\'on, Guanajuato; M\'exico}
\author{G. L\'opez Castro}
\email{glopez@fis.cinvestav.mx}
\author{L.-T. L\'opez Lozano}
\email{lao-tse@fis.cinvestav.mx}
\affiliation{Departamento de F\'{\i}sica, Cinvestav del IPN \\
Apartado Postal 14-740, 07000 M\'exico, D.F. M\'exico}
\bigskip

\bigskip

\bigskip

\begin{abstract}
The leading order contribution to the direct CP asymmetry in
$\tau^{\pm} \to K^{\pm}\pi^0 \nu_{\tau}$ decay rates is evaluated
within the Standard Model. The weak phase required for CP violation
is introduced through an interesting mechanism involving second
order weak interactions, which is also responsible for tiny
violations of the $\Delta S=  \Delta Q$ rule in $K_{l3}$ decays. The
calculated CP asymmetry turns out  to be of order $10^{-12}$,
leaving a large window for studying effects of non-standard sources
of CP violation in this observable.

\end{abstract}

\maketitle
\bigskip

\bigskip

Experimental searches for CP violating asymmetries in tau
lepton semileptonic decays have been carried out in the $\tau \to \pi\pi
\nu_{\tau}$  \cite{pipi} and  $\tau \to K_s  \pi \nu_{\tau}$ \cite{cleo} modes.
 Motivation for these searches in the context of beyond the Standard Model
approaches were provided in refs. \cite{cptau,tsai}. In ref. \cite{cleo},
the missing evidence for a non-zero CP asymmetry was interpreted in terms
of a (CP-violating) coupling $\Lambda$ due to a
charged scalar exchange and the limit $-0.172 <Im (\Lambda)<0.067$ (at
90\% c.l.) has been derived. The CP-odd observable studied in \cite{cleo}
depends upon two variables of a particular kinematical distribution
of semileptonic tau  decays  as long as  this  effect is assumed to have
its origin in the interference of  scalar and  vector form factors.

   Motivated by these searches and the possibility of further
improvements at a super-B-factory and taus produced in $W$ and $Z$ decays
at the LHC, in the present paper we compute the leading order Standard
Model
(SM) contribution to the  CP decay rate asymmetry between the two $\tau^{\pm} \to K^{\pm}
\pi^0  \nu_{\tau}$ decay channels. In order to have a non-zero CP asymmetry at the level of the
decay rate,  one requires that the CP-odd terms arise from the interference of terms in the same
 (vector or  scalar) angular configuration of the $K\pi$ system. Although
the  leading contribution to this CP asymmetry is a second order
weak process, it is interesting to estimate its actual magnitude to
be sure that any eventual observation of CP violation in these tau decay
experiments will have its origin beyond the SM framework.

\begin{figure}
  \includegraphics[width=12cm]{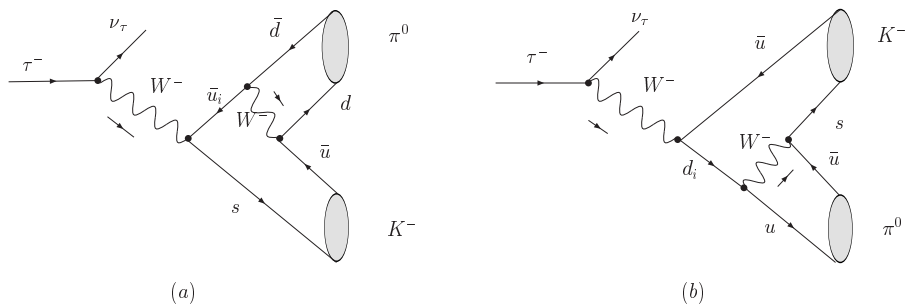}\\
  \caption{Higher order terms contributing to the $\tau^-  \to
K^-\pi^0 \nu_{\tau}$ decay.}\label{fig1a}
\end{figure}

   For the sake of clearness, first we keep our discussion as
general as possible. In the SM  the total amplitude for $\tau^-(p)
\rightarrow K^-(k)\pi^0(k')\nu_{\tau}(p')$ arising from tree-level
and the leading higher order terms in  Figures (\ref{fig1a}.a) and
(\ref{fig1a}.b) can be written in the following general form
\footnote{Note that the exchange of a charged scalar boson at the
tree-level such as the one advocated in ref. \cite{cleo} can be
absorbed into the definition of the effective form factor $F_0(t)$
if neutrino masses can be neglected.}: \be {\cal M} =
\frac{G_FV_{us}}{\sqrt{2}}
\left\{\bar{u}(p')\gamma^{\mu}(1-\gamma_5)u(p) F_+(t)
\left[(k-k')_{\mu}-\frac{\Delta^2}{t}q_{\mu} \right] +
\bar{u}(p')(1+\gamma_5)u(p) m_{\tau}
F_0(t)\frac{\Delta^2}{t}\right\}, \ee where $q=k+k'$ ($t=q^2$) is
the momentum transfer to the hadronic system, $\Delta^2   \equiv
m_K^2-m_{\pi}^2$ and $F_{+,0}(t)$ are the {\it effective} form
factors describing the hadronic matrix elements.

   The effective  vector and scalar form
factors can be written as follows:
\ba
F_+(t)&=&f_+(t)+a(t)\ , \\
F_0(t)&=&f_0(t)+b(t) \ , \ea where $f_i(t)$ are the usual tree-level
contributions and $a(t), \ b(t)$ denote the higher order terms
arising from Fig. (\ref{fig1a}) (as we will see later, Fig.
(\ref{fig1a}.b) does not induce a CP-violating phase). For the
purposes of numerical estimates of the CP asymmetry, we will choose
a simple model where the form factors at the tree-level are
dominated by a single vector or scalar strange resonance as follows
\cite{godina}:
\begin{equation}
 f_i(t)=\frac{f_i(0)m_i^2}{m_i^2-t-im_i\Gamma_i},
\ \ \ i=+,\ 0 ,
 \end{equation}
where ($m_i,\ \Gamma_i$) denote the mass and width of the resonance
in the
 corresponding vector or scalar configuration (respectively the $K^*(892)$ or $K^*_0(1430)$).
  Thus, the strong phase
corresponding to the
 tree-level amplitudes is determined by the decay width of these
resonances,  while the weak phase  is absent at the tree-level.

   The decay rate for the processes under consideration can be written in
the following form  \cite{godina}:
\be
\Gamma(\tau \to K\pi \nu) =
\frac{G_F^2m_{\tau}^5}{768\pi^3}|V_{us}|^2 I
\ee
where
\be
I=\frac{1}{m_{\tau}^6}\int_{(m_K+m_{\pi})^2}^{m_{\tau}^2}\frac{dt}{t^3}
(m_{\tau}^2-t)^2 \left[|F_+(t)|^2\left(1+\frac{2t}{m_{\tau}^2}\right)
\lambda^{3/2}(t,m_K^2,m_{\pi}^2) +3|F_0(t)|^2\Delta^4
\lambda^{1/2}(t,m_K^2,m_{\pi}^2)\right]\ ,
 \ee
and $\lambda(x,y,z)=x^2+y^2+z^2-2xy-2xz-2yz$.

Note that the effective scalar and vector form factors do not interfere at the level
of the hadronic spectrum or in the  integrated  rate
because they correspond to two
different angular  momentum configurations ($l=0,$ and 1) of the hadronic
system. This is interesting because the different strong and weak phases
required for CP violation should be present in the same angular
configuration of the two amplitudes. This mechanism is different from the
one used in ref.
\cite{cleo}, where the CP asymmetry vanishes at the level of the total
integrated rate and the hadronic spectrum, but it survives at the
level of a double kinematical distribution. On another hand, since the
contribution due
to the scalar form factor is  suppressed by powers of
the SU(3) breaking $\Delta^2$ parameter, one expects that the
dominant contribution to the CP asymmetry in the SM is given by the
interference of the vector form factors in Eq. (2).

Next, we proceed to evaluate the amplitudes corresponding to the
diagrams in Figs. (\ref{fig1a}). The weak vertices include the
couplings of fermions  to $W$ and would-be Goldstone bosons, and we
compute our amplitudes in the `t Hooft-Feynman gauge $\xi=1$. Note
that Fig. (\ref{fig1a}.a) (respectively (\ref{fig1a}.b)) involves
the exchange of an intermediate up-type quark $u_i=u,\ c,\ t$
(down-type quark $d_i=d,\ s,\ b$ ). Note also that the amplitudes
corresponding to Fig. (\ref{fig1a}.b) are proportional to
$|V_{ud_i}|^2$ and will not induce a (CP-violating) weak phase in
$a(t)$ or in $b(t)$.

   The hadronic matrix elements that correspond to Fig. (\ref{fig1a})
can be
evaluated in a similar way as perturbative  QCD techniques  allow to
evaluate  hadronic matrix elements relevant to $B$ decays \cite{du}.
 We will use the distribution amplitudes for pseudoscalar mesons
of momentum $p$ and mass $m_P$ \cite{du}
\be
\Psi_P(x,p)=\frac{-iI_C}{\sqrt{2N_C}}\phi_P(x)(\not p+m_P)\gamma_5\ ,
\ee
where $I_C$ is the identity in color space, $N_C$ is the number of colors
and $f_P$ the pseudoscalar decay constant ($f_{\pi}=130.7$ MeV and
$f_K=159.8$ MeV) \cite{pdg}. The  wavefunction of the pseudoscalar meson
is given by  $\phi_P(x)=\sqrt{3/2}f_Px(1-x)$ \cite{du}.

The form factors arising from the four diagrams in Fig.
(\ref{fig1a})(both, $W^-$ gauge bosons and $\phi^-$ would-be
Goldstones are understood in wavy-lines)  are given by:

\begin{eqnarray}
a(t)&=&-\frac{G_F}{\sqrt{2}}\frac{V_{ud}}{V_{us}}f_Kf_{\pi}\sum_{u_i=u,c,t}
V^*_{u_id}V_{u_is} \left[\left(
(t-m_{\pi}^2)(2+\frac{m_{u_i}m_u}{m_W^2})+\frac{m_d
m_{u_i}}{m_W^2}m_Km_{\pi}\right)I_i^1 \right.  \nonumber \\
&& + \left. \left(\frac{m_{u_i}
m_u}{m_W^2}m_{\pi}(m_{\pi}+m_d)+2m_{\pi}^2 -m_K(m_{\pi}+m_d)
\frac{m_{u_i}m_d}{m_W^2}\right)I_i^0\right]
\end{eqnarray}

\begin{eqnarray}
b(t)&=&-\frac{1}{\sqrt{2}}G_F\frac{V_{ud}}{V_{us}}f_Kf_{\pi}\sum_{u_i=u,c,t}
V^*_{u_id}V_{u_is} \left[\left(
(t-m_{\pi}^2-2m_K^2)(2+\frac{m_{u_i}m_u}{m_W^2})+\frac{m_d
m_{u_i}}{m_W^2}m_Km_{\pi}\right)I_i^1\frac{t}{\Delta^2}
\right.  \nonumber \\
&& +  \left(\frac{m_{u_i} m_u}{m_W^2}m_{\pi}(m_{\pi}+m_d)+2m_{\pi}^2
+m_K(m_{\pi}+m_d)
\frac{m_{u_i}m_d}{m_W^2}\right)I_i^0\frac{t}{\Delta^2}+a(t)
\nonumber \\
&&+\left(
\frac{m_Km_{\pi}^2}{m_W^2}(4m_{u_i}+\frac{m_d^3-m_um_{u_i}^2}{m_W^2})\right.
\nonumber
\\
&&
\left.-\frac{m_s}{2m_W^2}(t-m_K^2-m_{\pi}^2)(4m_{u_i}+\frac{m_d}{m_s}
\frac{m_u(m_d+m_{u_i})(m_{\pi}+m_{u_i})}{m_W^2})+\frac{m_{u_i}^2}{m_W^4}m_dm_{\pi}m_K(m_u+m_d)
\right)I_i^0 \nonumber \\
&& \left.+ \left( \frac{m_K}{2m_W^2}(t-m_K^2-m_{\pi}^2)
(4m_{u_i}+\frac{(m_d^3-m_um_{u_i}^2)}{m_W^2})
+m_K^2\frac{m_um_d}{m_W^4 } m_{\pi}(m_d+m_{u_i})\right)I_i^1 \right]
\end{eqnarray}
where we have defined the integral functions ($n=0,\ 1$):
 \begin{equation}
  I_i^n\equiv \int_0^1 dx
\frac{x^{n+1}(1-x)}{x^2m_K^2+x(t-m_K^2-m_{\pi}^2)+m_{\pi}^2-m_{u_i}^2+i\epsilon}
 \end{equation}
 Using the unitarity of the CKM mixing matrix, one gets
 \begin{eqnarray}
 \sum_{u_i=u,c,t} V^*_{u_id}V_{u_is}I_i^n&=&
 V^*_{ud}V_{us}(I_u^n-I_c^n)+V^*_{td}V_{ts}(I_t^n-I_c^n) \nonumber
 \\
 & \approx & V^*_{ud}V_{us}(I_u^n-I_c^n)- V^*_{td}V_{ts}I_c^n
 \end{eqnarray}
 where the last line  is obtained using the fact that $m_t$ is much
 larger than any mass of the particles involved in the process.
  The first term in last equation, being proportional to
$V^*_{ud}V_{us}$, will only
 contribute to the total rate but not to the $CP$ asymmetry. As a
 consequence, the $CP$ asymmetry will  only depend on the $I^n_c$
functions. In the kinematic region
 allowed for $t$, $I^n_c$ has a pole when the intermediary $c$ quark
is produced on its mass-shell in Fig.
 (\ref{fig1a}). Since the decay width  $\Gamma_c$ of the charm quark is
much smaller than its mass, it is
 possible to treat this pole through the $i\epsilon $ prescription
 for the quark propagator \cite{pole}. Integrating the $I^{0,1}_c$
functions, one
 gets
 \begin{eqnarray}
 I_c^0&=&\frac{1}{m_K^2}\left\{ \theta(m_c^2-t)
 \left[-1+\frac{(x^{+}-1)x^{+}}{x^{+}-x^{-}}\ln(\frac{x^{+}-1}{x^{+}})+\frac{(x^{-}-1)x^{-}}{x^{+}-x^{-}}\ln(\frac{x^{-}-1}{x^{-}})\right]
 \right. \nonumber \\
&&\left. -i\pi \theta(t-m_c^2)\frac{x^{+}(1-x^{+})}{|x^{+}-x^{-}|}
\right\} \\
I_c^1&=&\frac{1}{m_K^2}\left\{ \theta(m_c^2-t)
\left(\frac{1}{2}-(x^{+}+x^{-})+\frac{x^{+2}(x^{+}-1)}{x^{+}-x^{-}}
\ln(\frac{x^{+}-1}{x^{+}})+\frac{x^{-2}(x^{-}-1)}{x^{+}-x^{-}}
\ln(\frac{x^{-}-1}{x^{-}})\right)\right. \nonumber \\
&& \left.-i\pi
\theta(t-m_c^2)\frac{x^{+2}(1-x^{+})}{|x^{+}-x^{-}|} \right\}
\end{eqnarray}
where $ x^{\pm}=\left(-(t-m_K^2-m_{\pi}^2)\pm
\sqrt{\lambda(t,m_K^2,m_{\pi}^2)+4m_K^2m_c^2}\right)/(2m_K^2)$.

 The following remarks are important.  A simple inspection of Fig.
(\ref{fig1a}) tell us that the strong phases in $f_+(t)$
(respectively $f_0(t)$)
and $a(t)$ (respectively $b(t)$) are necessarily different. Indeed,
the higher order CP-violating contributions with $\bar{c}s$ and $\bar{t}s$
intermediate states can not produce the same resonance as the tree-level
contribution does in the $\bar{u}s$ channel.
Secondly, the
presence of the pole in the $I_c^n$ function
will produce a $CP$ conserving phase which could interfere with the
tree-level contribution. Thus, the tree-level and higher order
contributions have different weak and strong phases and will induce
a direct violation of CP.
 On another hand, note that the higher
order contributions $b(t)$ to the scalar form factors are suppressed
compared to $a(t)$ contribution as it will only interfere with $f_0$
form factor which is itself suppressed compared to $f_+$
\cite{godina}. So, we can safely neglect its contribution to the CP
asymmetry.

   The CP asymmetry can be written as follows:
\begin{eqnarray}
A_{CP}&=&\frac{\Gamma(\tau^+ \to
K^+\pi^0\bar{\nu}_{\tau})-\Gamma(\tau^- \to
K^-\pi^0\nu_{\tau})}{\Gamma(\tau^+ \to
 K^+\pi^0\bar{\nu}_{\tau})+\Gamma(\tau^-
\to K^-\pi^0\nu_{\tau})} \\
&\approx &
-\frac{\sqrt{2}G_F^3m_{\tau}^5\mbox{Im}(V_{us}V_{ud}^*V_{td}V_{ts}^*)f_Kf_{\pi}}{
768 \pi^3\Gamma (\tau^+ \to K^+\pi^0 \bar{\nu}_{\tau})}\times
I_{CP}\ ,
\end{eqnarray}
where we have neglected the $F_0$ contribution to
the $CP$ asymmetry and have kept the dominant contributions
from the amplitudes,
\begin{equation}
I_{CP}=\frac{1}{m_{\tau}^6}\int_{(m_K+m_{\pi})^2}^{m_{\tau}^2}\frac{dt}{t^3}
(m_{\tau}^2-t)^2 h(t) \left(1+\frac{2t}{m_{\tau}^2}\right)
\lambda^{3/2}(t,m_K^2,m_{\pi}^2)
\end{equation}
where $h(t)$ receives two dominant contributions according to the
values of $t$ in its kinematical domain :
\begin{eqnarray}
h(t)&=& \frac{2f_+(0)m_*^2}{(m_*^2-t)^2+m_*^2\Gamma_*^2} \left\{m_*
\Gamma_* \left(
(t-m_{\pi}^2)Re[I_c^1]+m_{\pi}^2Re[I_c^0]\right) \right.\nonumber \\
&&\left.-(m_*^2-t)\left(
(t-m_{\pi}^2)Im[I_c^1]+m_{\pi}^2Im[I_c^0]\right) \right\} \ ,
\end{eqnarray}
where $(m_*,\Gamma_*)$ denote the mass and width of the $K^*(892)$
resonance.
 Using the approximate expression for the
lifetime of the $\tau$ lepton $\tau^{-1} \approx 5
G_F^2m_{\tau}^5/192\pi^3$, one gets
\begin{eqnarray}
A_{CP}&\approx-
&\frac{\sqrt{2}G_F\mbox{Im}(V_{us}V_{ud}^*V_{td}V_{ts}^*)f_Kf_{\pi}}{20
B(\tau^+ \to K^+\pi^0 \bar{\nu}_{\tau})} \times I_{CP},
\end{eqnarray}
with $B(\tau^+ \to K^+\pi^0 \bar{\nu}_{\tau}) =(4.5 \pm 0.3) \times
10^{-3}$ \cite{pdg}.
 As we should have expected, the CP asymmetry becomes proportional
to the invariant measure of CP violation:
$J=\mbox{Im}(V_{us}V_{ud}^*V_{td}V_{ts}^*) = (2.88 \pm 0.33)\times
10^{-5}$ \cite{pdg}

\begin{figure}
  \includegraphics[width=10cm]{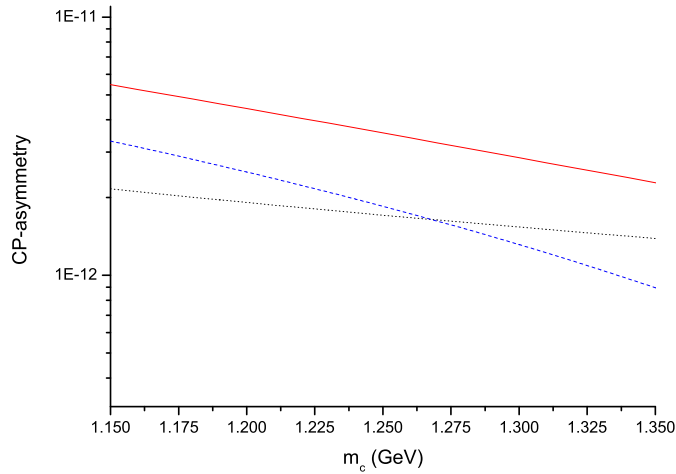}\\
  \caption{Contributions to the absolute value of the CP rate asymmetry as a function of the
charm quark mass $m_c$: contributions coming from the  pole of the
$c$ quark propagator (dotted-line) and coming from the interference
with the strong phase of the $K^{*}(892)$ (dashed-line), and their
sum (solid-line).}\label{fig2}
\end{figure}

  If we insert the expressions for the form factors  and  use
$f_+(0)=0.982/\sqrt{2}$ \cite{gl} and $m_c= 1.35$ GeV, we obtain the
following estimate for the decay rate asymmetry: \be
 |A_{CP} |\approx 2.3 \times 10^{-12} \ .
 \ee
In Figure (\ref{fig2}) we have plotted the absolute value of the CP
asymmetry coming from the  pole of the $c$ quark propagator
(dotted-line) and coming from the interference with the strong phase
of the $K^{*}(892)$ (dashed-line) as a function of the charm quark
mass within the $m_c$ range recommended in ref. \cite{pdg}. The
total absolute value of the CP asymmetry is represented by the
solid-line. As it can be observed from Fig. 2, the CP asymmetry is
not strongly sensitive on the value chosen for the charm quark mass,
and the contribution from the pole term becomes smaller for larger
values of $m_c$.

 The calculated CP rate asymmetry is small as expected from a CP asymmetry
that is generated by a second order weak interaction process. Thus, we can
conclude that this decay mode opens a large window to study constraints
on CP violation of a non-standard origin.

\begin{figure}
  \includegraphics[width=8cm]{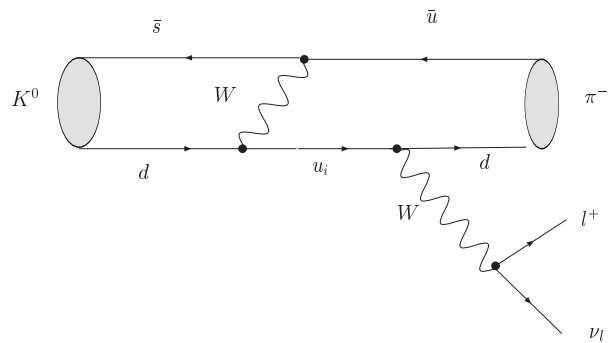}\\
  \caption{Higher order contribution to the $\Delta S = \Delta Q$ violating
$K^0 \to \pi^-l^+\nu_l$  decay.}\label{fig3}
\end{figure}

The  mechanism we have discussed here  to generate the  CP asymmetry
in the SM is unusual in the sense that the weak phase do not arise
from loop effects. The same mechanism can also generate a direct CP
decay rate  asymmetry for the isospin related $\tau^{\pm} \to K_S
\pi^{\pm} \nu_{\tau}$ decays.
 Also, such a mechanism could induce tiny
violations of the $\Delta S=\Delta Q$ rule. As is well known, this
selection rule implies that only the  decays $K^0 \to \pi^+l^-
\bar{\nu}_l$ and $\overline{K}^0 \to  \pi^- l^+ \nu_l$ are allowed
in the SM at the tree-level. The second order processes shown in Fig
(\ref{fig3}) would induce a very small $\Delta S=-\Delta Q$
component opening the possibility that the $K^0 \to \pi^-l^+\nu_l$
and $\overline{K}^0 \to  \pi^+ l^- \bar{\nu}_l$ can be allowed in
the SM. Present experimental limits on  $\Delta Q=\Delta S$
violating interactions in three-body semileptonic decays of $K_L$
are at the $10^{-3}$ level,  while the four-body $K^+ \to \pi^+\pi^+
e^- \bar{\nu}_e$ branching ratio has been bounded at the $10^{-8}$ level
\cite{pdg}.

\

\

\section{Acknowledgements}
The work of D.D. was supported by project PROMEP/103.5/04/1335
(M\'exico). G. L. C wants to acknowledge partial financial support
from Conacyt. D.D. wants to thank the hospitality of G.L.C  at Cinvestav
del IPN, where part of this work was done.

\


\begin{thebibliography}{40}
\bibitem{pole}
The appearance of this pole is intrinsic to the PQCD method, as it can be
seen from eq. (10). Notice however that at the hadron level a real meson
state with $c\bar{u}$ content (cutting both intermediate quark lines in
fig. 1a) is kinematically forbidden.
\end{thebibliography}
\end{document}